\documentclass[fleqn,10pt]{wlscirep}
\usepackage[utf8]{inputenc}
\usepackage[T1]{fontenc}
\usepackage{lineno}
\usepackage{amsfonts}
\usepackage{amsmath}
\usepackage{amssymb}
\usepackage{braket}
\usepackage{caption}
\usepackage{subcaption}
\usepackage[none]{hyphenat}
\usepackage{float}

\title{Generalising quantum imaginary time evolution to solve linear partial differential equations.}

\author[1,*]{Swagat Kumar}
\author[1,*]{Colin Michael Wilmott}
\affil[1]{Nottingham Trent University, Department of Physics and Mathematics, Nottingham, NG11 8NS, UK}
 \affil[*]{swagat.kumar02@ntu.ac.uk, colin.wilmott@ntu.ac.uk}

\begin{abstract}
The quantum imaginary time evolution (QITE) methodology was developed to overcome a critical issue as regards non-unitarity in the implementation of imaginary time evolution on a quantum computer. QITE has since been used to approximate ground states of various physical systems. In this paper, we demonstrate a practical application of QITE as a quantum numerical solver for linear partial differential equations. Our algorithm takes inspiration from QITE in that the quantum state follows the same normalised trajectory in both algorithms. However, it is our QITE methodology's  ability to track the scale of the state vector over time that allows our algorithm to solve differential equations. We demonstrate our methodology with numerical simulations and use it to solve the heat equation in one and two dimensions using six and ten qubits, respectively.
\end{abstract}
\begin{document}

\flushbottom
\maketitle
\thispagestyle{empty}
\section*{Introduction}

Quantum computers promise a paradigm shift for computing technology through their capability to solve problems that are  inaccessible with classical computers. It is well-understood that classical computers struggle to efficiently solve a class of problems known as optimisation, but a  principal promise of quantum computing relates to the significant improvements they bestow on the computational time needed to  solve such problems.
Quantum computers can be applied to a range of optimisation  problems that have widespread  value  in the real world, including   scheduling and planning\cite{Ho2004},  biochemical and computational biology\cite{Emani2021}, and financial risk\cite{Markowitz52}. Quantum computers were, however, originally proposed as a means to efficiently simulate quantum Hamiltonian dynamics\cite{Feynman1982}. Hamiltonian simulation is a task contained in the BQP complexity class\cite{Berstein1997}, which is significant because tasks in this class are believed to be intractable using classical computers. Many algorithms have been proposed for Hamiltonian simulation\cite{Lloyd96,Berry2007,Berry15,BerryChilds15,Low17, Haah_2021}, and  research has continued to improve the technique and  performance of such  simulations.    

Gate-based quantum computers solve the Hamitonian simulation problem by decomposing the unitary evolution operator into a series of smaller unitaries called quantum gates. The fundamental challenge is  how to instruct the quantum computer on the gate set needed to approximate the unitary.
Interestingly, the process to  undertake Hamiltonian simulation has not yet found an established method of choice. It is achieved by either directly implementing the time evolution operator or by determining the eigenspectrum of the Hamiltonian. State-of-the-art approaches for the former include quantum walks\cite{MADHU2023e13416}, qubitisation\cite{Low2019} and quantum signal processing\cite{Low17}. The latter is an equally difficult approach, and includes the variational quantum eigensolver (VQE)\cite{Peruzzo2014, McClean_2016,  Grimsley2019} and quantum imaginary time evolution (QITE)\cite{McArdle2019, Motta_2019}. These algorithms approximate the ground state of the Hamiltonian and can be applied recursively to obtain the complete eigenspectrum.

Computing the ground state energy of a Hamiltonian is of immense importance, such as in the computation of molecular and material energies \cite{Aspuru-Guzik05,Kandala2017,Kandala2019}, and wavefunctions\cite{LEHTOVAARA2007148, Kraus_2010,C5RA23047K,SHI2018245}.  Imaginary time evolution (ITE) is a successful classical method for determining the ground state of a Hamiltonian. 
By treating time as an imaginary number, the non-unitary time evolution operator generated by ITE
does not represent a physical process.
Classically solving the imaginary time Schr\"odinger equation  inherits the same computational complexities seen in classical simulations of quantum systems, namely the exponential overhead in maintaining the state of the system. This difficulty inspired the development of QITE as a technique to simulate  imaginary time evolution on a quantum computer by evolving a quantum state in imaginary time. In the ideal case, QITE guarantees convergence to the ground state, and, indeed, it  is a promising approach as attested by the increasing interest in its potential applications.

As a technique for pursuing the ground state of a Hamiltonian,  QITE may be implemented in two ways.  Variational QITE\cite{McArdle2019} is a hybrid quantum-classical algorithm that considers a system of differential equations linking to gradients of ansatz parameters in imaginary time and coefficients that depend on measurements of the ansatz, both of which contribute to  an update rule for finding the ground state.  Variational QITE is well suited for noisy intermediate-scale quantum (NISQ) devices as it has a fixed cost ansatz circuit.
Conversely, the imaginary time evolution may  move out of the ansatz space, implying that convergence may not reflect the true ground state. Furthermore,
designing a universal ansatz with a gate count that scales polynomially with the number of qubits is  a challenge, explaining  why most ansatzes are tailored to the Hamiltonian.

On the other hand, simulated QITE\cite{Motta_2019} outlines a quantum approach for simulating imaginary time evolution, by approximating the time evolution operator with Trotter products.
This implementation of simulated QITE  requires significantly fewer total measurements as compared to the  VQE algorithms to achieve the same level of convergence.
Simulated QITE with sufficiently large unitary domains does not suffer  barren plateaus, as is the case in the   variational approach.  In contrast, simulated QITE generates circuit depth increases that grow linearly with each imaginary time step.

Research to date has reported on ITE and QITE solely as methodologies for producing the ground state of a system where, under such implementations, information on the state vector's direction represents the key ingredient in determining the ground state. But research has not focused on fully understanding the significance of the role by the state vector's trajectory  during this evolution.   Indeed, in ITE and QITE implementations to date, the norm of the state vector is an irrelevant quantity in the computation. This is evident by how QITE simulates normalised time evolution but discards any information as to how the state vector scales in each time step. In this paper, we extend the scope of the QITE framework. 
Our extension offers a new avenue for exploration by  enlarging  the computational reach of the QITE methodology by our algorithm's ability to  track  the trajectory and scale of the state vector over time. We demonstrate how to approximate solutions to linear partial differential equations (PDEs) discretised via finite differences. We also  demonstrate our QITE methodology via numerical simulations and use it to   solve  the heat equation in one and two dimensions.

\section*{Preliminaries} 
The time evolution of a quantum state, $\psi(\vec{x},t)$, is governed by the Schr\"odinger equation which takes the form
\begin{equation}
    i\frac{\partial{\psi(\vec{x},t)}}{\partial t} = \hat{H} {\psi(\vec{x},t)},
    \label{eqn:schrodinger}
\end{equation}
where $\hat{H}$ is a Hermitian linear differential  operator known as the Hamiltonian. 
Since  the Hamiltonian is a Hermitian operator, it possesses a spectral decomposition  with eigenvalues $\lambda_n$ and corresponding normalised eigenstates $ {\psi_n}$. The lowest energy is known as the ground state of  the  system. Expanding the quantum state  $\psi(\vec{x},t)$ at the initial value $t = 0$  in terms of its  energy eigenstates, we have it that ${\psi(\vec{x},0)} = \sum_{n}{c_n{\psi_n(\vec{x})}}$, where $c_n$ denotes the overlap of ${\psi(\vec{x},0)}$ and ${\psi_n(\vec{x})}$. The quantum state at a later time $t$ is given by ${\psi(\vec{x},t)} = \sum_{n}{c_ne^{-\lambda_nit}{\psi_n}(\vec{x})}$. 
Applying the variable change $\beta = it$ to Eq.~(\ref{eqn:schrodinger}) yields the imaginary time Schr\"odinger equation  \begin{equation}\hbar\frac{\partial{\psi(\vec{x},\beta)}}{\partial \beta} = -\hat{H} {\psi(\vec{x},\beta)}.  \label{eqn:imag-schrodinger}
\end{equation} Since the Hamiltonian $\hat{H}$ remains the same, its solution takes the form  ${\psi(\vec{x},\beta=it)} = \sum_{n}{c_ne^{-\lambda_nt} {\psi_n}(\vec{x})}$. The state $ {\psi(\vec{x},\beta)}$   represents an exponentially decaying superposition of eigenstates,  which, in the limit of $\beta$ large, yields   ${\psi(\vec{x},\beta)} = {c_0e^{-\lambda_0\beta} {\psi_0}(\vec{x})}$. This demonstrates that ${\psi(\vec{x},\beta)}$  evolves parallel to the ground state of the system in the limit that imaginary time goes to infinity.

The QITE algorithm simulates the imaginary time evolution of quantum states via the Trotter product approximation. If we express the Hamiltonian as a linear combination of smaller, non-commuting operators $\hat{H} = \sum_I \hat{h}_I$, we can approximate the imaginary time evolution operator for a small time step of $\Delta t$ as
\begin{equation}
    e^{-\hat{H}\Delta t} = e^{-\sum_I\hat{h}_I\Delta t} 
    = \prod_I e^{-\hat{h}_I\Delta t} + O(\Delta t^2).
\end{equation}
Since each Trotter step, $e^{-\hat{h}_I\Delta t}$, in the product is non-unitary, the QITE algorithm approximates the normalised action of these operators on a unit quantum state $\ket{\bar{\psi}(t)}$ through a unitary operator $e^{-i\hat{A}\Delta t}$ such that
\begin{equation}
    e^{-i\hat{A}\Delta t}\ket{\bar{\psi}(t)} \approx \frac{e^{-\hat{h}_I\Delta t}\ket{\bar{\psi}(t)}}{\sqrt{\braket{\bar{\psi}(t)|e^{-2\hat{h}_I\Delta t}|\bar{\psi}(t)}}}.
    \label{eqn:normalised-qite-step}
\end{equation}
The support of $\hat{A}$ depends on the correlation length, $C$, of the state $\ket{\bar{\psi}(t)}$, which consists of $D = O(C)$ adjacent qubits surrounding the qubits in the support of $\hat{h}_I$\cite{Motta_2019}. Here, $D$ denotes the domain size of the approximating unitary. For an inexact implementation of the QITE algorithm, we choose $D$ according to the computational resources available. Our numerical implementations are based on the inexact QITE algorithm for $D=2,4,6$. For the best approximation, $\hat{A}$ should act on all qubits. However, calculating this operator requires solving a large system of linear equations that depend on the number of qubits the support of $\hat{A}$. To ease this computation, we can instead express the unitary as a product of unitaries $e^{-i\hat{A}_k\Delta t}$ such that the support of each $\hat{A}_k$ includes at most $D$ adjacent qubits.

\section*{Simulating PDEs with QITE}
QITE was introduced to replicate the imaginary time evolution of an initial state at every time step with the aim of producing the ground state solution. We propose to extend the scope of QITE by reimagining the role of the system's Hamiltonian $\hat{H}$ beyond its  immediate physical interpretation to a differential operator defining a  family of linear PDEs spanned by different choices of $\hat{H}$.  For instance, by considering the Hamiltonian $\hat{H}$ to be proportional to the Laplace operator $\nabla^2$, the imaginary time Schr\"odinger equation can be interpreted as the heat equation given by
\begin{equation}
    \frac{\partial {f(\vec{x},t)}}{\partial t} = \alpha\nabla^2 {f(\vec{x},t)}.
    \label{eqn:heat-eqn}
\end{equation}
The heat equation is the quintessential parabolic partial differential equation that has played a fundamental role in developing broader understandings of PDEs. The equation ranks amongst the most widely investigated topics in the physical sciences. The heat equation bridges to probability theory through its connection with the study of random walks and Brownian motion via the Fokker–Planck equation\cite{Risken1996}. The Black–Scholes equation\cite{BS73} of financial mathematics can be seen as a variant of the heat equation, and the Schr\"odinger equation reduces to a heat equation in imaginary time. From this position, QITE, offers an appealing route for simulating the normalised dynamics of this family of PDEs.

QITE seeks to determine the ground state of the system, where information relevant to the solution state is extracted from the state by taking measurement on the final quantum state. In practice, it should be expected that  the number of distinct measurements required to extract this relevant information will  scale polynomially with the number of qubits. For instance, in the case of the natural sciences, we  typically observe associated Hamiltonians having a polynomial number of non-zero terms in the Pauli basis. Determining the exact quantum state requires quantum state tomography and exponentially many measurements. However, if we restrict ourselves to simulating non-negative functions, we can reconstruct the quantum state using only the probability distribution of the Pauli Z, computational basis, measurements, since   the amplitudes of the quantum state are the square roots of the measurement probabilities of each computational basis state. 
For this reason, we will simulate the heat equation in one and two dimensions for non-negative solutions. To achieve this, and establish QITE as a methodology for simulating PDEs, we require, firstly, to discretise the system and, secondly, to encode the Hamiltonian in the Pauli basis. 

\subsection*{Discretising space} 
Propagation of a quantum state as determined  by the Schr\"odinger equation, Eq.~(\ref{eqn:schrodinger}), is  defined on a domain of continuous space. To simulate these dynamics with a discrete set of qubits, we are required to discretise the continuous wavefunction $\psi(\vec{x},t)$ to a discrete qubit state vector $\ket{\bar{\psi}(t)}$, and calculate the corresponding qubit Hamiltonian. We encode the continuous linear differential operator to a finite difference matrix. 
We will  first consider the one dimensional case before generalising  to higher dimensions.  

\subsubsection*{One dimensional space}
Let us consider a function defined on a one dimensional space domain $f:[a,b) \rightarrow \mathbb{C}$. This function can be encoded  into the state vector of $n$-qubits by storing $N=2^n$ uniformly spaced samples of the function in an unnormalised state vector
\begin{equation}
    \ket{f} = \sum_{k=0}^{N-1} f\left(a + kh\right) \ket{k} = \sum_{k=0}^{N-1} f_k \ket{k},
    \label{eqn:state-encoding}
\end{equation}
with the spacing $h = \frac{b-a}{N}$ and $\ket{k}$ denoting elements of standard basis. Next, let us consider   approximating  a linear partial differential operator on the discretised space using the method of finite difference approximation of derivatives.  A first order finite difference takes the form $f(x+b)-f(x+a)$ and is classified as the central difference when we have $\delta^1_h[f](x) = f(x+h/2) - f(x-h/2)$, for spacing $h$. Higher order partial differential operators are approximated by the central finite differences given by
\begin{eqnarray}
    \frac{\partial^m f(x)}{\partial x^m} \approx \frac{\delta_h^m[f] (x)}{h^m}
    \ \textrm{where} \ \delta_h^m[f] (x) = \sum_{i=0}^m (-1)^i \binom{m}{i} f\left(x + \left(\frac{m}{2} - i\right)h\right).
\end{eqnarray}
Of particular interest is the   second order partial  differential  operator
that appears in the heat equation, which can be approximated by  the difference operator  
\begin{equation}
    \hat{\delta}_{h}^{2} \ket{f} = \frac{1}{h^2} \sum_{k=0}^{N-1} (f_{k+1} - 2f_k + f_{k-1}) \ket{k}.
\end{equation}
The boundary conditions determine the values of $f_{-1}=f(a-h)$ and $f_N=f(b)$. The difference operator under the zero boundary conditions,  $f_{-1} = f_N = 0$, is represented by the following matrix written in the standard basis
\begin{equation}
    \frac{1}{h^2}
    \begin{pmatrix}
    -2 &  1 &        &        &    \\
     1 & -2 &  1     &        &     \\
       &  1 & \ddots & \ddots &     \\
       &    & \ddots & \ddots &   1 \\
       &    &        & 1      &  -2 
\end{pmatrix}.
\label{eqn:zero-bc-matrix}
\end{equation}
The difference operator under periodic boundary conditions, $f_{-1}=f_{N-1}$ and $f_N=f_0$, is represented by the following matrix written in the standard basis
\begin{equation}
    \frac{1}{h^2}
    \begin{pmatrix}
    -2 &  1 &        &        &   1\\
     1 & -2 &  1     &        &     \\
       &  1 & \ddots & \ddots &     \\
       &    & \ddots & \ddots &   1 \\
    1 &    &        & 1      &  -2 
\end{pmatrix}.
\label{eqn:periodic-matrix}
\end{equation}
Let $\hat{D}^{(n)}_0$ denote the $n$-qubit second-order finite difference Hamiltonian under zero boundary conditions such that 
$\hat{D}^{(n)}_0 = h^2 \hat{\delta}_{h}^{2}.$ 
It then follows that, in the Pauli operator basis,
\begin{equation}
\hat{D}_0^{(1)} = 
\begin{pmatrix}
    -2 & 1 \\
    1 & -2
    \end{pmatrix}
= {-2\hat{I} + \hat{X}}
.
\label{eqn:one-qubit-D}
\end{equation}
To determine the Pauli basis representation of $n$-qubit  second-order finite difference Hamiltonian under zero  boundary conditions, we  define 
\begin{eqnarray}
   \hat{A}^{(n)}_{\swarrow} := \hat{A}^{(1)}_{\swarrow} \otimes \hat{A}^{(n-1)}_{\swarrow} \quad \rm{where} \quad   \hat{A}^{(1)}_{\swarrow} := \frac{\hat{X} - i\hat{Y}}{2} =
    \begin{pmatrix}
    0 & 0 \\
    1 & 0
    \end{pmatrix}, 
    \label{eqn:A-south-west}
\end{eqnarray}
and 
\begin{eqnarray}
        \hat{A}^{(n)}_{\nearrow} := \hat{A}^{(1)}_{\nearrow} \otimes \hat{A}^{(n-1)}_{\nearrow}\quad \rm{where} \quad      
    \hat{A}^{(1)}_{\nearrow} :=
    \frac{\hat{X} + i\hat{Y}}{2} =
    \begin{pmatrix}
    0 & 1 \\
    0 & 0
    \end{pmatrix}.
    \label{eqn:A-north-east}
\end{eqnarray}
The two-qubit Hamiltonian is given as
\begin{equation}
    \hat{D}^{(2)}_0 = 
   \left(\begin{array}{cc|cc}
       -2 &  1 & 0 & 0 \\
        1 & -2 & 1 & 0 \\ \hline
        0 & 1  &-2 & 1 \\
        0 & 0  & 1 &-2
   \end{array}\right).
\end{equation}
Using Eqs.~(\ref{eqn:one-qubit-D}-\ref{eqn:A-north-east}), we have 
\begin{equation}
    \hat{D}^{(2)}_0 = 
     \begin{pmatrix}
        \hat{D}^{(1)}_0 & \hat{A}^{(1)}_{\swarrow} \vspace{5pt}\\
        \hat{A}^{(1)}_{\nearrow} & \hat{D}^{(1)}_0
    \end{pmatrix}
    = \hat{I} \otimes \hat{D}^{(1)}_0
    + \hat{A}^{(1)}_{\swarrow} \otimes \hat{A}^{(1)}_{\nearrow}
    + \hat{A}^{(1)}_{\nearrow} \otimes \hat{A}^{(1)}_{\swarrow}.
\end{equation}
From Eq.~(\ref{eqn:zero-bc-matrix}), we can show that the $n$-qubit Hamiltonian $\hat{D}^{(n)}_0$ has the form 
\begin{equation}
    \hat{D}^{(n)}_0 
    = \begin{pmatrix}
        \hat{D}^{(n-1)}_0 & \hat{A}^{(n-1)}_{\swarrow} \vspace{5pt}\\
        \hat{A}^{(n-1)}_{\nearrow} & \hat{D}^{(n-1)}_0
    \end{pmatrix}
    = \hat{I} \otimes \hat{D}^{(n-1)}_0
    + {\hat{A}^{(1)}_{\swarrow}} \otimes \hat{A}^{(n-1)}_{\nearrow}
    + {\hat{A}^{(1)}_{\nearrow}} \otimes \hat{A}^{(n-1)}_{\swarrow}.
\end{equation}
Similarly, we define $\hat{D}^{(n)}_p$ to be the
$n$-qubit second-order finite difference Hamiltonian under periodic boundary conditions. It can be shown from Eq.~(\ref{eqn:periodic-matrix}) that $\hat{D}^{(n)}_p$ takes the form
\begin{equation}
    \hat{D}^{(n)}_p = \hat{D}^{(n)}_0 + \hat{A}^{(n)}_{\swarrow} + \hat{A}^{(n)}_{\nearrow}.
\end{equation}

\subsubsection*{Higher dimensional space} Generalising the state encoding for the one dimensional case to higher space dimensions is achieved by taking the tensor product of the qubit registers of the associated dimensions. For example,  a function defined on a two dimensional space domain    
$f : [a_1,b_1)\times[a_2,b_2) \rightarrow \mathbb{C}$ can be encoded with $2n$ qubits by storing $N^2$ samples in the unnormalised state vector
\begin{equation}
    \ket{f} = \sum_{k_1, k_2 = 0}^{N-1}  f(a_1 + k_1h_1,\ a_2 + k_2 h_2) \ket{k_1}\ket{k_2} = \sum_{k_1, k_2=0}^{N-1} f_{k_1,\ k_2} \ket{k_1}\ket{k_2}
\end{equation}
with spacings $h_i = \frac{b_i - a_i}{N}$ for $i=1,2$. Similarly,  we can  construct the associated finite difference operator by taking the tensor products of the underlying one dimensional operators. For instance, the two dimensional Laplace operator, $\nabla^2 = \partial^2_x + \partial^2_y$, is represented by the following $2n$-qubit finite difference operator

\begin{equation}
    \hat{L}^{(n)}_{h_1, h_2}\ket{f} = \left[\frac{\hat{D}^{(n)}}{h_1^2} \otimes \hat{I}^{\otimes n} 
    + \hat{I}^{\otimes n} \otimes \frac{\hat{D}^{(n)}}{h_2^2}\right] \ket{f} 
    = \sum_{k_1,k_2 = 0}^{N-1}
    \left(\frac{f_{k_1+1,\ k_2} - 2f_{k_1,\ k_2} + f_{k_1-1,\ k_2}}{h_1^2}
    + \frac{f_{k_1,\ k_2+1} - 2f_{k_1,\ k_2} + f_{k_1,\ k_2-1}}{h_2^2}\right)\ket{k_1}\ket{k_2}.
    \label{eqn:laplace-hamiltonian}
\end{equation}

\subsection*{Obtaining solutions from the state vector}
Although QTIE simulates the trajectory of the PDE solution, it does not account for how the length of the state vector changes over time. To achieve our intended application, we must also approximate the norm at each time step and rescale the state vectors obtained from QITE to match the complete dynamics of the PDE solution.

\subsubsection*{Measuring the state vector}
If we know that the original function only takes on non-negative values in the region we are solving for, the state vector $\ket{f}$ will only have non-negative amplitudes in the computational basis. 
We will, therefore, restrict ourselves to PDEs involving only even-ordered differential operators as they are Hermitian and represented by real matrices in the computational basis. This  ensures that  the quantum state $\ket{f}$ will not contain any phase information and can be reconstructed by taking the square root of its computational basis measurement probability distribution. 

\subsubsection*{Reconstructing the norm}
In the reconstruction of the norm, we seek to approximate
the squared norm, $c(t) = \braket{\bar{\psi}(t)|e^{-2\hat{h}\Delta t}|\bar{\psi}(t)}$, of the non-unitary evolution operator $e^{-\hat{h}\Delta t}$ at each simulated time step of size $\Delta t$. 
QITE achieves this using the linear order approximation ${c^\prime(t) = 1 + 2\Delta t\braket{\bar{\psi}(t)|\hat{h}|\bar{\psi}(t)}}$. Assuming that a QITE implementation of $N_T$ time steps has perfect fidelity, we can express the vector containing samples of the PDE solution, $f(x,t=N_T\Delta t)$, as
\begin{equation}
    \ket{f(N_T\Delta t)} = \|\ket{f(0)}\|\left(\prod_{j=0}^{N_T-1} \sqrt{c(j\Delta t)}\right) \ket{\bar{\psi}(N_T\Delta t)}.
\end{equation}
The theoretical squared norm   at the $k$-th time step is given as the product of the $k$ individual  squared norms 
\begin{equation}
    C_f(k\Delta t) = \prod_{j=0}^{k-1} c(j\Delta t).
\end{equation}
An approach to   approximate $C_f(t)$ would be to consider the  product of the linear approximants
\begin{equation}
    C^\prime(k\Delta t) = \prod_{j=0}^{k-1} c^\prime(j\Delta t).
\end{equation}
The issue with this approach is that the relative errors associated to  $c^\prime(j\Delta t)$, for $j =0,\cdots,k-1$,  compound in the product, which leads to a significant deviation from the theoretical norm with each additional time step. 
To mitigate the accumulation of errors in the running product, $C^\prime(k\Delta t)$,   we undertake a strategy to rescale the norm after every $K$ time steps.
To implement this strategy, we require the    normalised ground state of the Hamiltonian, $\ket{\bar{\psi}_0}$, and its associated eigenvalue, $\lambda_0$.
Let $C_*(t)$ denote a good approximation for the squared norm. Under the assumption that our QITE simulation has  high fidelity, that is,  ${\ket{f(t)} \approx \|\ket{f(0)}\|\sqrt{C_*(t)} \ket{\bar{\psi}(t)}}$, we  then have it that  
\begin{eqnarray}\label{approx}
     {\frac{\braket{\bar{\psi}_0|f(t)}}{\|\ket{f(0)}\|}}
        \approx \sqrt{C_*(t)}{\braket{\bar{\psi}_0|\bar{\psi}(t)}}.   
        \end{eqnarray}
Using $\ket{f(t)} = \sum_{k=0}^{N-1}{e^{-\lambda_k t}} \ket{\bar{\psi}_k}\bra{\bar{\psi}_k}\ket{f(0)}$, it follows that 
\begin{eqnarray}    
\begin{aligned}
    {\frac{\braket{\bar{\psi}_0|f(t)}}{\|\ket{f(0)}\|}} &= \frac{\bra{\bar{\psi}_0}\left(\sum_{k=0}^{N-1}{e^{-\lambda_k t}} \ket{\bar{\psi}_k}\bra{\bar{\psi}_k}\right)\ket{f(0)}}{\|\ket{f(0)}\|}  \\
        &= \sum_{k=0}^{N-1}{e^{-\lambda_k t}}\braket{\bar{\psi}_0| \bar{\psi}_k}\bra{\bar{\psi}_k}\frac{\ket{f(0)}}{\|\ket{f(0)}\|}  \\ 
         &= \sum_{k=0}^{N-1}{e^{-\lambda_k t}}\delta_{k,0}\bra{\bar{\psi}_k}\frac{\ket{f(0)}}{\|\ket{f(0)}\|}  \\ 
         &= e^{-\lambda_0 t}\braket{\bar{\psi}_0|\bar{f}(0)}, 
       \end{aligned}
\end{eqnarray} 
and, consequently, Eq.~(\ref{approx}) may be rewritten as   
   \begin{eqnarray}  
   e^{-\lambda_0 t}{\braket{\bar{\psi}_0|\bar{f}(0)}}
         \approx
        {\sqrt{C_*(t)} \braket{\bar{\psi}_0|\bar{\psi}(t)}},\end{eqnarray}
 from which we deduce an  approximation for $C_*(t)$ as   \begin{eqnarray} 
    C_*(t) 
          \approx 
        e^{-2\lambda_0 t}\frac{\braket{\bar{\psi}_0|\bar{f}(0)}^2}{\braket{\bar{\psi}_0|\bar{\psi}(t)}^2} .
        \label{eqn:norm-correction}
\end{eqnarray}
To mitigate the error propagation in $C^\prime(t)$, we periodically rescale the running product to  $C_*(t)$ after every $K$ time steps, giving us an improved approximation of the squared norm;
\begin{equation}
    C_\psi(k\Delta t) = \begin{cases}
       C_*(k\Delta t) & \text{for } k \equiv 0\ \text{mod}\ K \\
        C_\psi((k-1)\Delta t) \cdot c^\prime(k\Delta t) & \text{for } k \not\equiv 0\ \text{mod}\ K
    \end{cases}.
\end{equation}

\begin{figure}[H]
    \includegraphics[width=\textwidth]{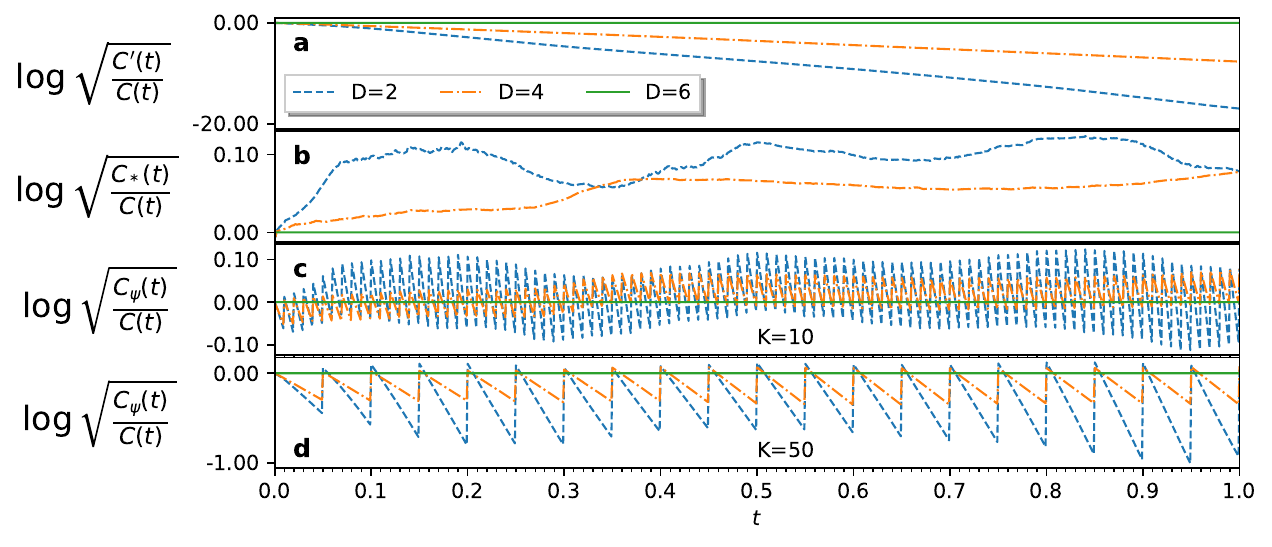}
    \caption{\textbf{Comparison of different norm reconstruction strategies.} The graphs show the base-10 logarithms of the ratios of the reconstructed norms to the analytical norm for a 6-qubit square wave evolution against the heat equation with $\alpha=0.8$ over 1,000 simulated time steps.
    \textbf{(a)} We see how the errors in $c^\prime$ compound in the their product $C^\prime$.
    \textbf{(b)} The fidelity of our QITE simulation increases with $D$, and we see how $C_*$ is a better approximation for the norm at higher fidelities.
    \textbf{(c)} $C_\psi$ combines information from $c^\prime$ and $C_*$ and is, on average, a better approximation of the norm for sufficiently small $K$. 
    \textbf{(d)} At higher values of $K$, the errors in $c^\prime$ are able to compound, making the approximation worse on average.
    }
    \label{fig:K-effect}
\end{figure}

\subsection*{Comparing QITE with analytical evolution} 

Our methodology performs a unitary approximation of a linear PDE and  provides  an estimate on how the norm evolves. This information allows us to  obtain approximate solutions to the  PDE.
Let  $\ket{\bar{f}(t)} = \ket{f(t)}/\sqrt{C_f(t)}$ denote  the normalised state vector containing scaled samples of the analytical solution $f(x,t)$ of the PDE, and  $\ket{{\psi}(t)} =  \sqrt{C_\psi(t)} \ket{\bar{\psi}(t)}$   denote the rescaled state vector containing samples of the solution $\psi(x,t)$ approximated by our QITE methodology.
When restricted   to functions that only take on non-negative values, the fidelity of our  QITE implementation, given by
\begin{equation}
    F(t) = \braket{\bar{f}(t)|\bar{\psi}(t)},  
\end{equation}
measures the accuracy of our normalised evolution.
The ratio between the approximated and analytical norms is 
\begin{equation}
    r(t) = \sqrt{\frac{C_\psi(t)}{C_f(t)}},
\end{equation}
which measures the accuracy of our reconstruction.
For $N$ samples, we can write the mean squared error, MSE, as  
\begin{eqnarray}
\begin{aligned}
    \text{MSE}(t) &= \frac{\|\ket{f(t)} - \ket{\psi(t)}\|^2} {N}\\
    &= \frac{C_f(t) + C_\psi(t) - 2\sqrt{C_f(t) C_\psi(t)}\braket{\bar{f}(t)|\bar{\psi}(t)}}{N} \\
    &= \frac{C_f(t) + r^2(t) C_f(t) - 2 r(t) C_f(t) F(t)}{N} \\
    &= \frac{2C_f(t)}{N}\left[\frac{1+r^2(t)}{2} - r(t)F(t)\right].
\end{aligned}
\label{eqn:mse}
\end{eqnarray}
Equation~(\ref{eqn:mse}) demonstrates the mean squared error to be a useful metric because it correlates to both the fidelity and norm ratio of our approximation.

\section*{Results}
 To demonstrate how our QITE methodology can be used to solve PDEs, we target a simulation of the heat equation,  Eq.~(\ref{eqn:heat-eqn}).  When expressed in terms of the imaginary time Schr\"odinger equation, the   Hamiltonian operator corresponding to  the heat equation   is $\hat{H} = -\alpha\nabla^2$. We performed numerical simulations of our QITE methodology   for  domain sizes $D=2,4$ and $6$, from which we obtained approximate solutions to the heat equation for various initial states and boundary conditions. Across our experiments, we set  $\alpha=0.8$,    simulated the dynamics from $t=0$ to $t=1$, and used a  grid spacing of $h=0.1$, and a time step $\Delta t=0.001$. 
 The results reported in Fig.~\ref{fig:K-effect} allowed us to decide on a norm correction frequency of $K=10$, as the log norm ratio oscillated around zero for this choice of constants. 

 \begin{figure}
    \centering
    \includegraphics[width=\textwidth]{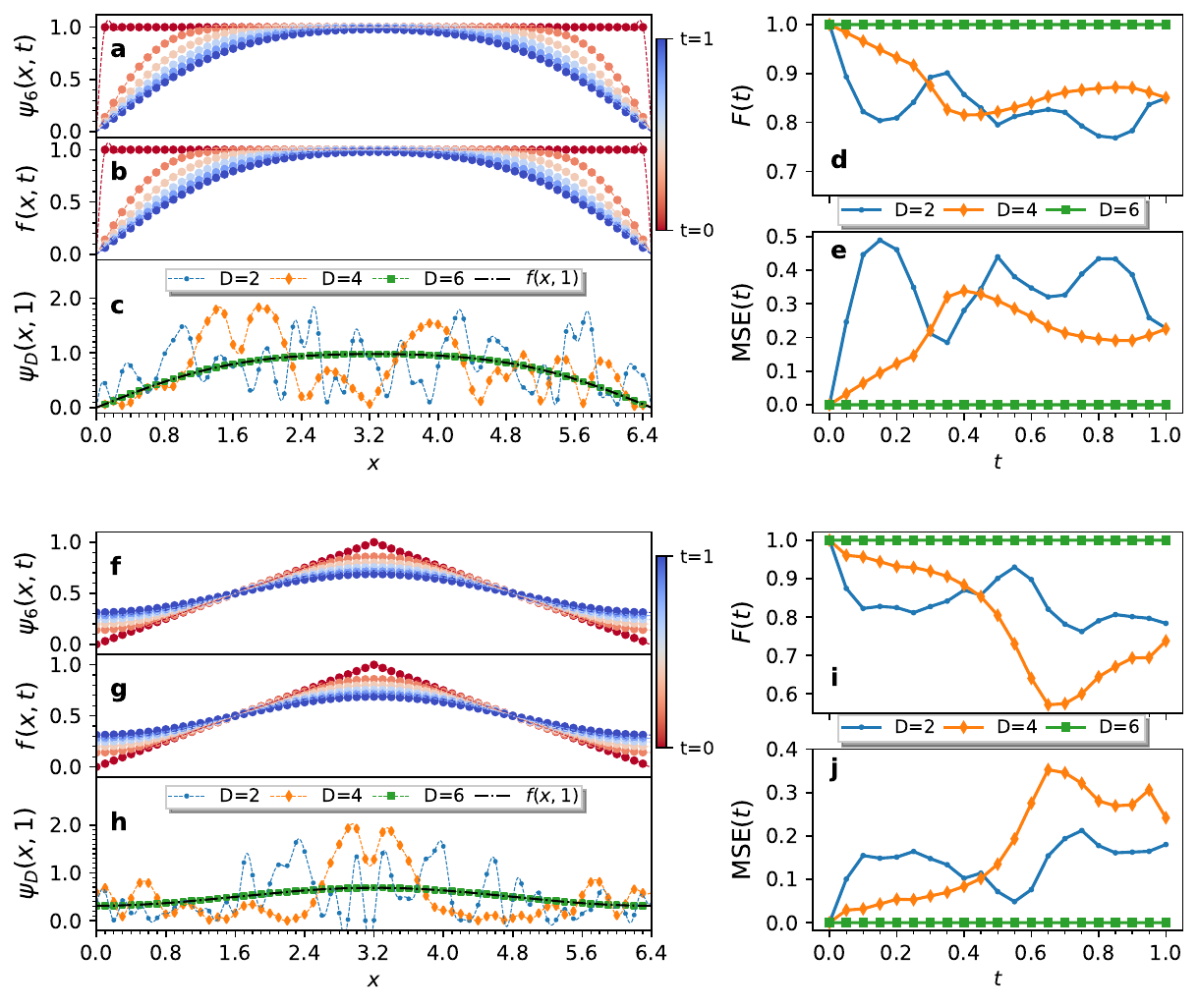}
    \caption{\textbf{QITE simulations for the heat equation in one spatial dimension.} The figure compares our QITE solutions for various domain sizes $D$, $\psi_D(x,t)$, to the analytical solutions, $f(x,t)$, of the heat equation with $\alpha=0.8$. The function samples were encoded in the state vector of 6 qubits. 
    We show results of two simulations; \textbf{(a)-(e)} refer to a square wave with zero boundary conditions, while \textbf{(f)-(j)} refer to a triangle wave with periodic boundary conditions.
    The norm was corrected at every $K=10$ simulated time steps.
    \textbf{(a)} and \textbf{(f)} show the solutions obtained from QITE for a domain size of $D=6$ at different times. \textbf{(b)} and \textbf{(g)} show the analytical solutions $f(x,t)$ at the same time steps, as indicated by the color of the curves. The dots indicate the function samples, connected by their Fourier interpolations. \textbf{(c)} and \textbf{(h)} compare the states produced by the $D=2,4,6$ QITE approximations at time $t=1$ to the corresponding analytical solution. \textbf{(d)} and \textbf{(i)} show the fidelity of the QITE approximations over time. \textbf{(e)} and \textbf{(j)} show the mean squared error of the QITE approximations to the analytical solutions over time.}
    \label{fig:1d-solutions}
\end{figure}
 
\begin{figure}
    \centering
    \includegraphics[width=0.76\textwidth]{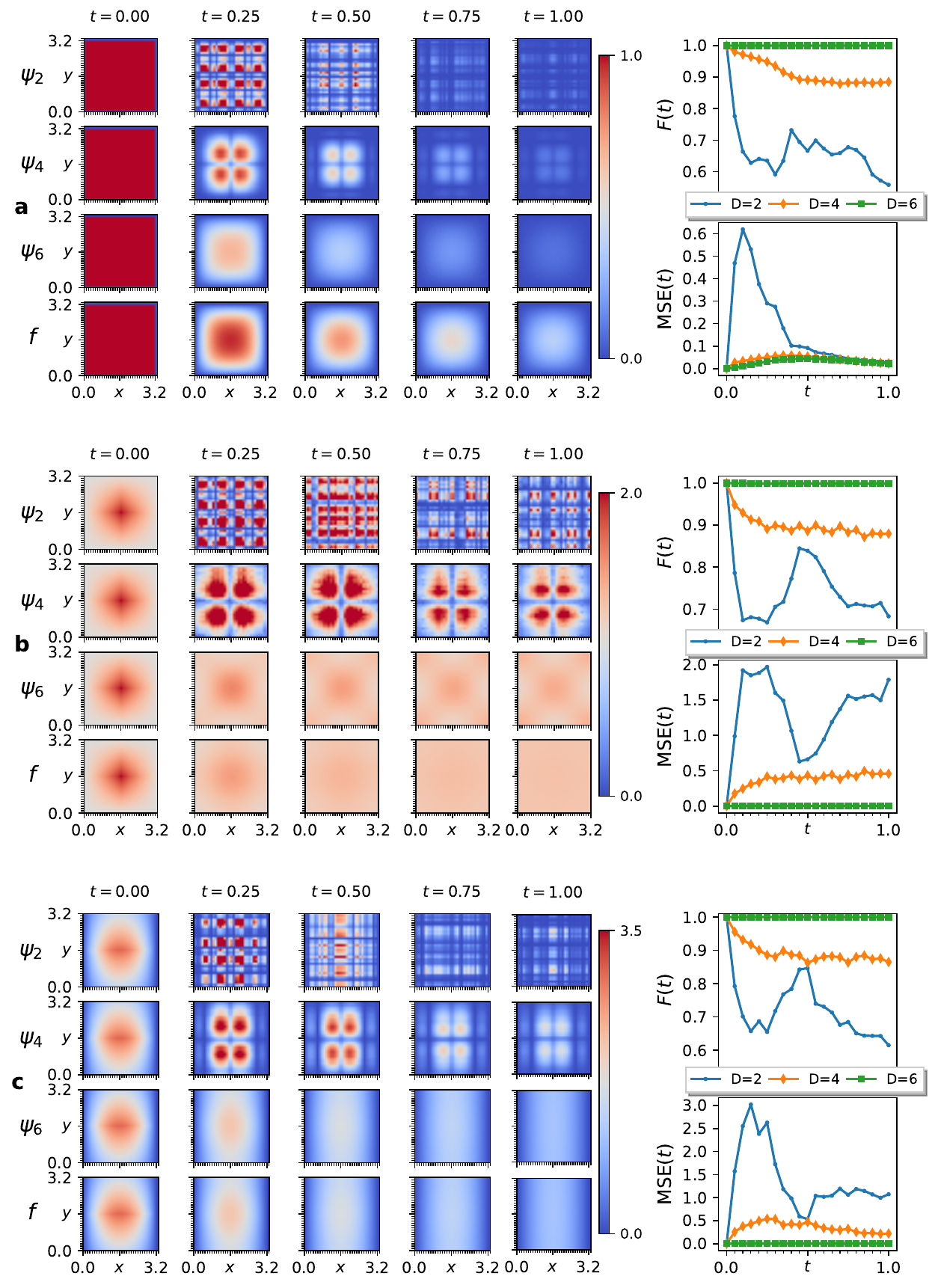}
    \caption{\textbf{QITE simulations for the heat equation in two spatial dimensions.} 
    The figure compares our QITE solutions with domain size $D$, $\psi_D(x,y,t)$, to  the analytical solutions, $f(x,y,t)$, of the two dimensional heat equation with diffusion coefficient $\alpha=0.8$. The function samples were encoded in the state vector of 10 qubits, with the $x$ and $y$ axes mapped to 5 qubits each. The norm was corrected at every $K=10$ time steps.
    The plots on the left show the QITE solutions corresponding to $D=2,4,6$ and the analytical solution (top to bottom) at different times (columns), with the color of pixels corresponding to the value of the function according to the color bar.
    The fidelity and mean squared error of the approximations are also displayed on the right.
    The initial states and boundary conditions for the experiments are as follows: in \textbf{(a)}, the initial state is a two dimensional square wave with zero boundary conditions in both $x$ and $y$ directions; in \textbf{(b)}, the initial state  is  a product of triangle waves in $x$ and $y$ with height 1, a total offset of 1, and periodic boundary conditions in both $x$ and $y$ directions; in  \textbf{(c)}, the initial state is a product of a triangle wave in $y$ with height 1, and an inverted parabola in $x$ with a maximum height of 1.5, and zero boundary conditions in the $x$ direction and periodic boundary conditions in the $y$ direction.
    }
    \label{fig:2d-heat-eqn-results}
\end{figure}

 Our implementation of the QITE methodology as a PDE solver supports two families  of boundary conditions, namely, the zero boundary conditions, $f(0)=f(L)=0$, and  periodic boundary conditions, $f(x)=f(x+L)$. Figure~\ref{fig:1d-solutions} demonstrates the results of our simulations for the one dimensional heat equation for the zero and periodic boundary conditions. We stored the solutions in the state vector of $n=6$ qubits, giving us the boundary lengths $L=6.5$ in the case of the zero boundary conditions, and $L=6.4$ in the case of the periodic boundary conditions. Since the $D=6$ approximation covers the entire set of interacting qubits, our QITE implementation demonstrated a perfect fidelity to the analytical solution and zero mean squared error. Figure~\ref{fig:2d-heat-eqn-results} demonstrates our solutions for the two dimensional heat equation, where we considered all combinations of zero and periodic boundary conditions in each spatial dimension. We used   10 qubits to store the function samples, distributing   $n=5$ sampling qubits for both $x$ and $y$ directions, which yielded  boundary lengths $L=3.3$ for the zero boundary conditions, and $L=3.2$ for periodic boundary conditions. Since the two dimensional Laplace Hamiltonian, shown in Eq.~(\ref{eqn:laplace-hamiltonian}), does not have interactions between the $x$ and $y$ axes' sampling qubits, we chose unitary domains to covered the five $x$ and $y$ qubits individually. This allowed the $D=6$ approximation to cover the entire set of interacting qubits, again yielding perfect fidelity to the analytical solutions. We also demonstrated a zero mean squared error Fig.~\ref{fig:2d-heat-eqn-results} (b) and (c). Note that in Fig.~\ref{fig:2d-heat-eqn-results} (a), there is a small deviation in mean squared error from zero due to compounding errors in approximating the norm.
 
\section*{Discussion}
As regards to the norm correction frequency, we empirically determined that a norm correction frequency $K=10$ was sufficient to approximate the dynamics of the heat equation for our choice of constants. Further investigation is needed to determine a logical correlation between the choice of constants and a suitable value for $K$. In the case of when norm correction is required but that   the    exact ground state is not known, we can estimate this state by running a QITE simulation over a long period of imaginary time to get approximations for the ground state,  $\ket{\bar{\Psi}_0} \approx \ket{\bar{\psi}_0}$ and its eigenvalue, $\Lambda_0 = \braket{\bar{\Psi}_0|\hat{H}|\bar{\Psi}_0} \approx \lambda_0$. Adopting  this state as a heuristic for the ground state, we can substitute $\Lambda_0$ and $\ket{\bar{\Psi}_0}$ in Eq.~(\ref{eqn:norm-correction}) to get an approximate norm correction factor. 
In relation to   the function encoding schemes, the finite difference matrices approximating the second derivative operator have at most three non-zero entries in each row. These entries  indicate the interaction of the basis states as they pertain to the determination of  the basis state amplitudes in the final state vector. In particular, the output amplitude of basis state $\ket{k}$ depends on the basis states $\ket{k-1}, \ket{k},$ and $\ket{k+1}$. Under our  encoding scheme, each basis state $\ket{k}$ is mapped to elements of the  computational basis state in lexicographical order.
Consequently, under the direct encoding scheme, approximations for $D<n$ are unable to capture all of the interactions in the finite difference matrix. This is most easily seen in the horizontal and vertical bands in the centre of the $D=4$ approximations in Fig.~\ref{fig:2d-heat-eqn-results}.
A different encoding scheme that allows us to capture all possible qubit interactions for $D<n$ should permit  perfect fidelity with an exponential speed-up compared to the direct encoding above. 
Finally, as regards to general boundary conditions, the QITE methodology examined here   only allows us to capture the zero and periodic boundary conditions, since these  conditions correspond to Hermitian finite difference matrices.

\section*{Conclusions}
In this work, we  demonstrated a  new and practical application of QITE  as a quantum numerical solver for linear PDEs. Our methodology adopts QITE's ability to model the normalised trajectory of a  quantum state. Additionally, our methodology also tracks the scale of the state vector over time. 
It is the interaction between these two features that 
has enabled us to  broaden the scope of QITE to 
approximate solutions to linear PDEs discretised via finite differences. 
Using numerical simulations, we implemented our methodology to solve the 
heat equation in one and two dimensions, using six and ten qubits, respectively. In our experiments, we demonstrated perfect fidelity along with a mean squared error converging to zero.
 
\section*{Code availability} 
The code that supports the findings of this study is  available from the authors upon reasonable request.

\bibliography{sources}

 \section*{Acknowledgements}

C.M.W. and S.K. gratefully acknowledge the financial support of the Engineering and Physical Sciences Research Council (EPSRC) through the Hub in Quantum Computing and Simulation (EP/T001062/1).

\section*{Author contributions}
C.M.W. conceived the project. S.K. performed the numerical simulations. C.M.W. and S.K. analyzed the data and interpreted the results, and proposed improvements. C.M.W. and S.K. both contributed to the writing and editing of the manuscript.  

\section*{Competing interests}
The authors declare no competing interests.

\end{document}